# Three-Dimensional Organic Microlasers with Low Lasing Thresholds Fabricated by Multiphoton and UV Lithography

Vincent W. Chen,[1†] Nina Sobeshchuk,[2,3†] Clément Lafargue,[2†] Eric S. Mansfield,[1] Jeannie Yom,[1] Luke Johnstone,[1] Joel M. Hales,[1] Stefan Bittner,[2] Séverin Charpignon,[2] David Ulbricht,[2] Joseph Lautru,[2] Igor Denisyuk,[3] Joseph Zyss,[2] Joseph W. Perry[1*] and Melanie Lebental[2*]

[1]*School of Chemistry and Biochemistry and Center for Organic Photonics and Electronics, Georgia Institute of Technology, Atlanta, GA 30332-0400*
[2]*Laboratoire de Photonique Quantique et Moléculaire, CNRS UMR 8537, Institut d'Alembert FR 3242, Ecole Normale Supérieure de Cachan, 61 Avenue du Président Wilson, F-94235 Cachan, France*
[3]*ITMO University, Kronverkskiy pr.49, 197101, Saint Petersburg, Russia*
[†] These authors contributed equally to this work
*[joe.perry@gatech.edu](mailto:joe.perry@gatech.edu), *[lebental@lpqm.ens-cachan.fr](mailto:lebental@lpqm.ens-cachan.fr)

**Abstract:** Cuboid-shaped organic microcavities containing a pyrromethene laser dye and supported upon a photonic crystal have been investigated as an approach to reducing the lasing threshold of the cavities. Multiphoton lithography facilitated fabrication of the cuboid cavities directly on the substrate or on the decoupling structure, while similar structures were fabricated on the substrate by UV lithography for comparison. Significant reduction of the lasing threshold by a factor of ~30 has been observed for cavities supported by the photonic crystal relative to those fabricated on the substrate. The lasing mode spectra of the cuboid microresonators provide strong evidence showing that the lasing modes are localized in the horizontal plane, with the shape of an inscribed diamond.

## References and links

1. H. Kogelnik, and C. V. Shank, "Stimulated emission in a periodic structure," Applied Physics Letters **18**, 152-154 (1971).
2. Á. Costela, I. Garcia-Moreno, and R. Sastre, "Solid-State Dye Lasers," in Tunable Laser Applications, F. J. Duarte, ed. (CRC Press, 2009).
3. S. Chénais, and S. Forget, "Recent advances in solid-state organic lasers," Polymer International **61**, 390-406 (2012).
4. L. He, Ş. K. Özdemir, and L. Yang, "Whispering gallery microcavity lasers," Laser & Photonics Reviews **7**, 60-82 (2013).
5. J. Zhu, Ş. K. Özdemir, Y.-F. Xiao, L. Li, L. He, D.-R. Chen, and L. Yang, "On-chip single nanoparticle detection and sizing by mode splitting in an ultrahigh-Q microresonator," Nat. Photon **4**, 46-49 (2010).
6. A. B. Matsko, Practical Applications of Microresonators in Optics and Photonics (CRC Press, Boca Raton, FL, 2009).
7. S. K. Y. Tang, R. Derda, Q. Quan, M. Lončar, and G. M. Whitesides, "Continuously tunable microdroplet-laser in a microfluidic channel," Opt. Express **19**, 2204-2215 (2011).
8. M. Kuwata-Gonokami, R. H. Jordan, A. Dodabalapur, H. E. Katz, M. L. Schilling, R. E. Slusher, and S. Ozawa, "Polymer microdisk and microring lasers," Opt. Lett. **20**, 2093-2095 (1995).
9. M. Pöllinger, D. O'Shea, F. Warken, and A. Rauschenbeutel, "Ultrahigh-Q Tunable Whispering-Gallery-Mode Microresonator," Physical Review Letters **103**, 053901 (2009).
10. D. K. Armani, T. J. Kippenberg, S. M. Spillane, and K. J. Vahala, "Ultra-high-Q toroid microcavity on a chip," Nature **421**, 925-928 (2003).
11. Y. F. Xiao, C. H. Dong, C. L. Zou, Z. F. Han, L. Yang, and G. C. Guo, "Low-threshold microlaser in a high-Q asymmetrical microcavity," Opt. Lett. **34**, 509-511 (2009).

12. E. Bogomolny, N. Djellali, R. Dubertrand, I. Gozhyk, M. Lebental, C. Schmit, C. Ulysse, and J. Zyss, "Trace formula for dielectric cavities. II. Regular, pseudointegrable, and chaotic examples," Physical Review E **83**, 036208 (2011).
13. G. D. Chern, A. W. Poon, R. K. Chang, T. Ben-Messaoud, O. Alloschery, E. Toussaere, J. Zyss, and S. Y. Kuo, "Direct evidence of open ray orbits in a square two-dimensional resonatorof dye-doped polymers," Opt. Lett. **29**, 1674-1676 (2004).
14. M. Lebental, N. Djellali, C. Arnaud, J. S. Lauret, J. Zyss, R. Dubertrand, C. Schmit, and E. Bogomolny, "Inferring periodic orbits from spectra of simply shaped microlasers," Physical Review A **76**, 023830 (2007).
15. S. Lozenko, N. Djellali, I. Gozhyk, C. Delezoide, J. Lautru, C. Ulysse, J. Zyss, and M. Lebental, "Enhancing performance of polymer-based microlasers by a pedestal geometry," Journal of Applied Physics **111**, 103116-103119 (2012).
16. C. Lafargue, S. Bittner, S. Lozenko, J. Lautru, J. Zyss, C. Ulysse, C. Cluzel, and M. Lebental, "Three-dimensional emission from organic Fabry-Perot microlasers," Applied Physics Letters **102**, 251120 (2013).
17. B. H. Cumpston, S. P. Ananthavel, S. Barlow, D. L. Dyer, J. E. Ehrlich, L. L. Erskine, A. A. Heikal, S. M. Kuebler, I. Y. S. Lee, D. McCord-Maughon, J. Qin, H. Rockel, M. Rumi, X.-L. Wu, S. R. Marder, and J. W. Perry, "Two-photon polymerization initiators for three-dimensional optical data storage and microfabrication," Nature **398**, 51-54 (1999).
18. S. Kawata, H.-B. Sun, T. Tanaka, and K. Takada, "Finer features for functional microdevices," Nature **412**, 697-698 (2001).
19. S. R. Marder, J.-L. Brédas, and J. W. Perry, "Materials for Multiphoton 3D Microfabrication," MRS Bulletin **32**, 561-565 (2007).
20. C. N. LaFratta, J. T. Fourkas, T. Baldacchini, and R. A. Farrer, "Multiphoton Fabrication," Angewandte Chemie International Edition **46**, 6238-6258 (2007).
21. G. von Freymann, A. Ledermann, M. Thiel, I. Staude, S. Essig, K. Busch, and M. Wegener, "Three-Dimensional Nanostructures for Photonics," Advanced Functional Materials **20**, 1038-1052 (2010).
22. A. d. Campo, and C. Greiner, "SU-8: a photoresist for high-aspect-ratio and 3D submicron lithography," Journal of Micromechanics and Microengineering **17**, R81 (2007).
23. H. Lorenz, M. Despont, N. Fahrni, N. LaBianca, P. Renaud, and P. Vettiger, "SU-8: a low-cost negative resist for MEMS," Journal of Micromechanics and Microengineering **7**, 121 (1997).
24. B. F. Howell, and M. G. Kuzyk, "Amplified spontaneous emission and recoverable photodegradation in polymer doped with Disperse Orange 11," J. Opt. Soc. Am. B **19**, 1790-1793 (2002).
25. I. Gozhyk, G. Clavier, R. Méallet-Renault, M. Dvorko, R. Pansu, J. F. Audibert, A. Brosseau, C. Lafargue, V. Tsvirkun, S. Lozenko, S. Forget, S. Chénais, C. Ulysse, J. Zyss, and M. Lebental, "Polarization properties of solid-state organic lasers," Physical Review A **86**, 043817 (2012).

## 1. Introduction

Since the earliest demonstration of stimulated emission in a Rhodamine 6G dye/gelatin thin-film, patterned with a distributed feedback grating [1], there has been interest in the development of organic microlasers (OMLs). Over the intervening decades a wide variety of microresonator designs and organic dye-based gain media have been reported [2-4]. In recent years, OMLs have attracted considerable interest as tunable light sources for sensing [5], and have shown potential as highly emissive taggants, optical security features and for on-chip photonic applications [6]. Various OML structures such as vertical-cavity surface-emitting lasers have utilized reflective Fabry-Perot cavities, whereas waveguides, microspheres [7], micro-rings [8], micro-bottle [9], micro-toroids [10], and asymmetric cavity [11] structures utilize total internal reflection, eliminating the need for cavity mirrors and enabling a route to simple integration with chip-scale photonic devices. In recent years, the fabrication of Fabry-Perot like resonators, microsquares and other polygonal structures, as well as stadium structures, have been demonstrated by means of traditional UV lithography. While the modes of whispering gallery type resonators are based on ray trajectories along the cavity circumference, the modes of polygonal resonators are often localized on various types of ray trajectories as revealed by the analysis of their spectra[12-14].

Typically two-dimensional structures have been investigated, which means that the thickness is much smaller than the other dimensions. Here we address the fabrication and properties of three-dimensional (3D) cavities, i.e. the dimensions of the resonator are comparable in all directions. One of the many advantages of multiphoton lithography (MPL) over traditional 2D patterning techniques is the ability to fabricate arbitrary 3D structures. This provides the unique opportunity to investigate various designs that could enable decoupling of the microresonators from the substrate. Physical contact of the microresonators with the underlying substrate generally leads to an increase of mode losses and diffraction at the resonator-substrate interface [15, 16].

In this paper, we report on the lasing properties of cuboid OMLs, fabricated through the use of 3D MPL [17-21] and conventional UV microlithography [22, 23], in order to examine OMLs that are distinctively three dimensional. First, we compare the characteristics of OMLs produced by the two fabrication methods. Then, we consider one of the interesting challenges associated with substrate supported microlasers, that is, the loss due to the small index mismatch between the dye-doped polymer gain material and a glass substrate, or the optical coupling of the emission into and absorption loss due to a semiconductor substrate. We then investigate approaches to decouple the lasing emission of cuboid OMLs from the substrate, in an effort to reduce the lasing threshold and increase the output power of such structures. Body-centered tetragonal (BCT) woodpile photonic-crystal (PC) pedestals, with approximately eight layers and having close to the same edge dimension as the cuboid, were used as decoupling structures. We have examined the lasing spectra of the cuboids as a function of edge length and height to understand the impact of the cuboid dimensions and PC pedestal structures on the lasing mode spectra and lasing threshold.

## 2. Experimental methods

### 2.1. Organic microlaser fabrication by MPL

For MPL fabrication, a photo-crosslinkable resin comprised of a propoxylated glyceryl triacrylate monomer (SR9020, Sartomer) and 0.5 wt% of a 2-benzyl-2-(dimethylamino)-4'-morpholinobutyrophenone (Aldrich) photoinitiator. 1,3,5,7,8-pentamethyl-2,6-di-t-butylpyrromethene-difluoroborate complex (PM-597, Exciton, see Fig. 1) was incorporated into the photocrosslinkable resin at a concentration of 0.8 wt% to form the gain medium. The pre-polymer resin was mixed on a stirring plate overnight prior to use. The sample cell was formed by sandwiching the pre-polymer resin between a glass microscope slide and a glass coverslip (VWR International) with a Kapton spacer (60 μm thickness). The microscope slides were pretreated with 3-(trimethoxysilyl)propyl methacrylate (Aldrich) to promote adhesion of the polymer to the glass substrate. Photocrosslinking was performed using 12-20 mW average power from a mode-locked Ti:sapphire laser (Tsunami, Spectra-Physics, 750 nm, ∼ 120 fs pulses, 82 MHz repetition rate) that was directed through a raster scanner (MRC1024ES, BioRad) and focused into the sample cell by a 40×ELWD microscope objective (Nikon, NA=0.6).

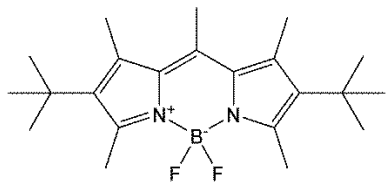

Fig. 1. Chemical structure of PM- 597

Cuboid OMLs with edge lengths ranging from 28 μm to 58 μm were fabricated by controlling the raster scanning area (i.e. layer area) via the zoom adjustment and pixel density. The heights (between 26 μm and 46 μm) of the OMLs were defined by computer controlled translation of the sample stage (MP-285, Sutter Instruments) along the optical axis. A fabrication protocol, via a home-built LabView program, involving a raster scan followed by a controlled step in the z-direction (step size of 1 μm) was implemented to achieve good overlap of adjacent layers (~4 µm thickness) and to ensure uniform exposure throughout the fabricated structure. The stage was synchronized with a mechanical shutter to expose each layer for a time duration of 4.3 s. The OMLs were fabricated directly on a glass substrate or on BCT woodpile PC pedestals with an in-plane (x,y) spacing of 5 µm and a layer spacing (z) of 2 µm. The latter were fabricated by computer-controlled translation of the sample with respect to a stationary focused laser beam. The structures were developed using 4-methyl-2-pentanone (Aldrich) with two washing cycles of 5 minutes each.

*2.2. Organic microlaser fabrication by UV lithography*

For the UV lithography fabrication procedure a commercially available SU8 photoresist with high viscosity (SU8 2025 by MicroChem) was used as the polymer matrix. It is well-suited for the production of thick microstructures with high aspect ratios. The resin was ultrasonicated and stirred after addition of PM-597 (0.5 wt% relative to the polymer in solution) to fully dissolve the laser dye prior to use. A film of the resin was deposited on either silicon oxide (2 µm) on silicon wafers (Si/SiO$_2$) or on glass slides by spin-coating. The film thickness was determined by the viscosity of the resist and the parameters of the spin-coating process. Thicknesses in the range of 20–100 µm were achieved. The samples were then subjected to a pre-exposure bake (5 min at 65°C and 60 min at 95°C). The length of the pre-bake was significantly increased relative to the standard process recommended by MicroChem to reduce the amount of residual solvent. The UV lithography process was implemented on a manual mask aligner (MJB4, Suss MicroTech). The duration of exposure depended on the film thickness and varied in the range of 20–100 seconds. Afterwards the samples were post-exposure baked (3 min at 65°C and 10-15 min at 95°C) and developed for 5–10 min in SU8 developer followed by an isopropanol wash.

*2.3. Scanning electron microscope images*

The OMLs were investigated using scanning electron microscopy (SEM, Hitachi S3400N or Zeiss Ultra60), optical microscopy (UMA600, Agilent), and a stylus profilometer (Dektak 3ST, Veeco) in order to determine the quality of the cavity surfaces and the perpendicularity and parallelism of the facets. The samples were coated by a 10 nm thick gold layer prior to SEM imaging, which was subsequently removed with potassium iodide. SEM images of OMLs of various shapes, including structures with rounded features, are presented in Fig. 2. They were analyzed using image analysis software (ImageJ, NIH) to obtain the widths, lengths, and heights of the OMLs. The SEM images demonstrate that cavities with high aspect ratios, very few defects, and almost arbitrary shapes can be obtained from laser dye-doped resins with both fabrication techniques. In the following sections, we will focus on cuboid OMLs.

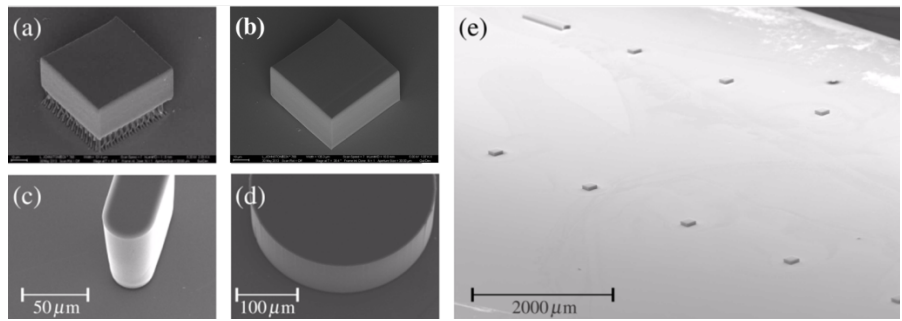

Fig. 2. SEM images of microlasers. (a) cuboid OML with PC pedestal structure fabricated by MPL; (b) cuboid OML fabricated on glass substrate by MPL; (c-e) OMLs fabricated by UV lithography.

*2.4. Optical characterization*

Organic microlasers were pumped at room temperature with a pulsed frequency-doubled Nd:YAG laser (532 nm, 500 ps, 10 Hz) through the top surface (*xy*-plane) of the OML, along the *z*-axis, as shown in Fig. 3(a). The energy of the pump laser was adjusted by the combination of a half-wave plate and a polarizer. An additional half-wave and quarter-wave plate controlled the polarization state of the pump laser [not shown in Fig. 3(a)]. A circularly polarized pump beam was used for the measurements reported here. The size of the pump spot was adjusted through use of lenses such that the pump spot covered a whole microcavity, but none of the neighboring ones, so that only one OML was pumped at a time. The full width at half maximum of the spot was about 200 μm. The light emitted by the OML in the *x*-*y*-plane was collected by a lens and transferred via a fiber to a spectrometer (Spectra Pro 2500i, Acton Research). The sample was positioned on a rotational stage to allow measurement of the emission into different directions in the *x*-*y*-plane. A second optical setup (not shown here) was also used to measure the emission out of the *x*-*y*-plane [16]. The spectra shown here were measured in the *x*-*y*-plane and in a direction perpendicular to a cavity side wall.

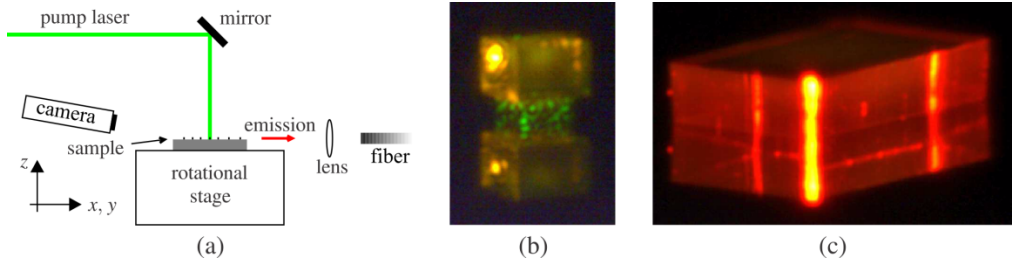

Fig. 3. (a) Schematic drawing of the experimental setup (not to scale). (b) optical image of a lasing cuboid OML with fabricated by MPL on a PC pedestal and (c) optical image of a cuboid OML on a Si/SiO$_2$ substrate fabricated by UV lithography. Note the duplication of the images due to reflection from the substrate.

A camera (UI324xCP-C, IDS Imaging) with a high-magnification zoom lens (Zoom 6000, Navitar) was used to obtain images of the cavities under lasing conditions. The camera was slightly inclined with respect to *x*-*y*-plane [Fig. 3(a)]. Images of two different lasing cuboid OMLs are shown in Fig. 3(b) and (c). Fig. 3(b) shows a cavity fabricated by MPL with a side length (*a*) of 36.0 μm and a height (*h*) of 34.7 μm with an underlying PC pedestal, while Fig. 3c shows a cavity fabricated by UV lithography on a Si/SiO$_2$ wafer with a side length of 200 μm and a height of 50 μm. The cavities were rotated such that the majority of the lasing emission (yellow), coming from a side facet, was in the direction of the camera. For the cuboid OML on the PC pedestal, some green pump light scattered by the PC pedestal and edges of the OML is also captured by the camera.

## 3. Results and discussion

*3.1. Spectral properties*

Fig. 4 shows a typical multimode lasing spectrum of a cuboid cavity (50(*a*)×50(*a*)×25(*h*) μm$^3$), that was fabricated by UV lithography, with pump intensity just above the threshold. The resonances show a linewidth of about 0.1 nm and are regularly spaced. Such a series of equidistant resonances suggests that the observed resonant modes are localized on a periodic ray trajectory inside the cavity (a so-called periodic orbit, PO) [12, 14]. If this is the case, then the resonance wavenumbers, $k_m$, fulfill a semiclassical quantization condition of the form:

$$nl_{po}k_m = 2\pi m + \varphi \qquad (1)$$

where $n$ is the refractive index of the cavity, $l_{po}$ is the length of the PO, $m$ is the mode quantum number, and $\varphi$ is a phase shift due to, e.g., reflections at the boundaries of the cavity. Hence, the free spectral range (FSR) $\Delta k = k_m - k_{m-1}$ of the family of modes is given by:

$$\Delta k = \frac{2\pi}{n_g l_{po}} \qquad (2)$$

where $n_g$ is the group refractive index taking into account the dispersion of the refractive index of the medium [14]. The Fourier Transform (FT) of the spectrum (i.e., the transformation from wave number $k$ to optical length $l_{opt}$) then exhibits a peak at $l_{opt} = n_g l_{po}$. Therefore, the FT of the lasing spectrum allows the identification of possible underlying POs. The FT of the lasing spectrum of the cuboid cavity is shown as an inset in Fig. 4. It exhibits a peak at $l_{opt}$ = 244 µm, as well as several harmonics. From ellipsometry measurements, the refractive index of SU-8 is about 1.6. Hence, the only PO consistent with the experimental optical length is a 2D diamond orbit in the x,y plane, which is totally internally reflected at each of the vertical side walls of the cavity as shown in the inset of Fig. 4. The optical length of the diamond mode is $l_{opt} = n_g 2\sqrt{2}a$ = 226 µm, where $a$ = 50 µm is the length of the sidewall, which is in good agreement with the optical length observed in the Fourier spectrum.

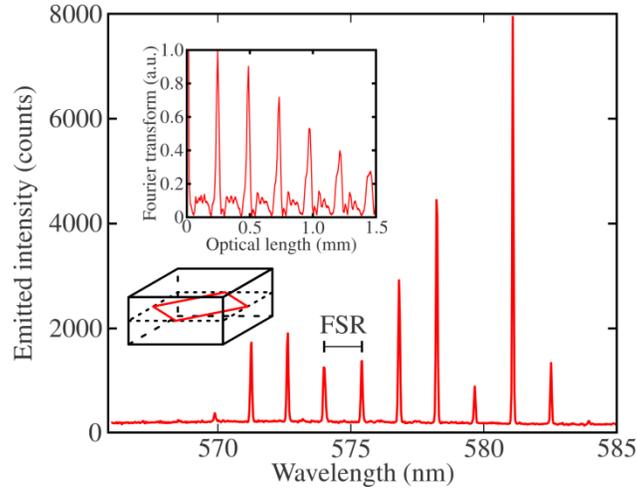

Fig. 4. Lasing spectrum of a cuboid cavity (50×50×25µm³) fabricated by UV lithography. The insets show the Fourier transform of the lasing spectrum and a sketch of the diamond periodic orbit.

The same general structure of the lasing mode spectrum was also observed for the other cuboid cavities. Fig. 5 shows the spectra of a cuboid cavity on a PC pedestal fabricated by MPL and pumped with two different intensities. Just above the threshold (upper curve), the spectrum shows primarily a single family of equidistant modes like in Fig. 4. It should be noted that the same spectrum (with varying amplitude) was observed in all emission directions in and out of the *x-y* plane. When the edge lengths of the cavities were varied the structure of equidistant resonances was preserved. However, as the side length of the OML decreased, the number of observed modes decreased as a result of the increase of the FSR. Fig. 6(a) shows the optical length corresponding to the FSR as a function of the edge length *a* for three cavities with an identical height of $h$ = 42 µm. The error bars correspond to the width of the peaks in the FTs of the lasing spectra. The optical length of the diamond orbit is indicated by the blue line, using a refractive index $n_g$ = 1.51, determined by prism coupling and ellipsometry. The good agreement between the blue line and the data points confirms that the observed families of modes are related to the diamond orbit. Furthermore, Fig. 6(b) shows the optical lengths obtained from the FTs of the lasing spectra for a series of cavities with varying height but identical side length $a$ = 58 μm. This shows that the optical length is independent of the cavity height within the error margins. This is a clear indication that the underlying PO is two-dimensional, i.e., it rests within the *x*-*y*-plane and has no reflections at the top and bottom facets. The optical length predicted for the diamond orbit is indicated by the blue line and agrees well with the measured optical lengths. All these results give evidence that the observed lasing modes of the cuboid cavities are related to the diamond orbit.

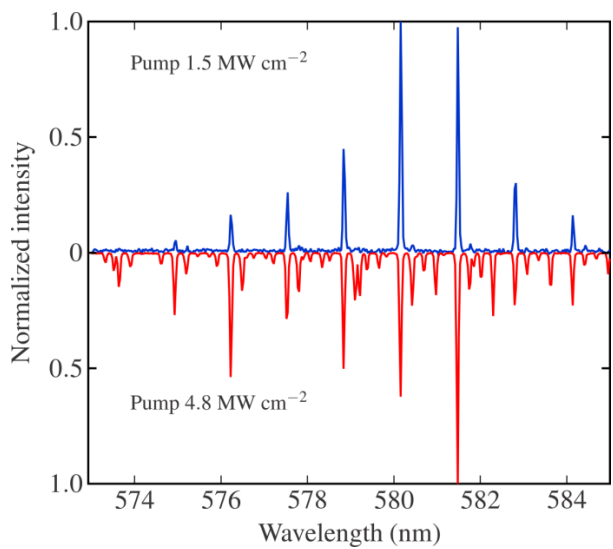

Fig. 5. Normalized lasing spectra of a cuboid cavity (58×58×32μm³) on a PC pedestal fabricated by MPL pumped at two different intensities.

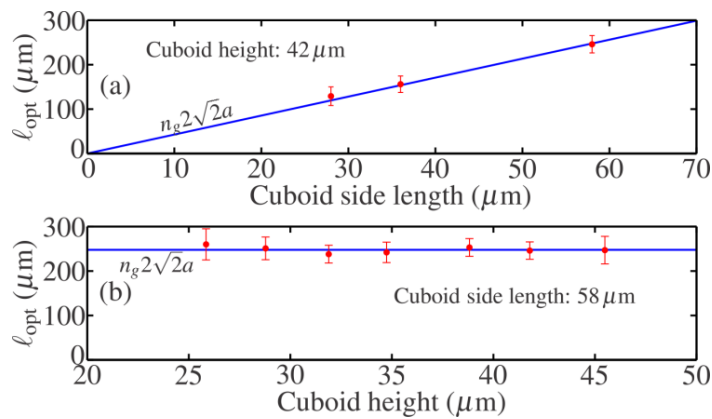

Fig. 6. (a) Optical length *versus* cavity side length $a$ for cuboid cavities with a height of $h$ = 42 μm, fabricated by MPL. The blue line is the optical length of the diamond orbit for a group refractive index of $n_g$ = 1.51. (b) Optical length *versus* the height of cavities with side length $a$ = 58 μm. The blue line is the optical wavelength of the diamond orbit for a group refractive index of $n_g$ = 1.51.

### 3.2. Photostability

The photostability of the 3D OMLs was investigated since the longevity of the laser dye incorporated into a polymer matrix is critical for practical applications. Firstly, the cuboid

OML samples show no evidence of degradation during storage at room temperature and ambient atmosphere over a period of months. When pumped optically, the lifetime of the dye depends on the pump intensity. Recovery phenomena are observed as well if the OML is not pumped during a sufficiently long time (typically one hour) [24]. The photostability curve of a cuboid cavity supported by a PC pedestal fabricated by MPL is shown in Fig. 7. The OML was pumped just above its threshold without interruption for approximately 100,000 pump pulses and the emitted intensity showed only a very small variation for this number of pulses.

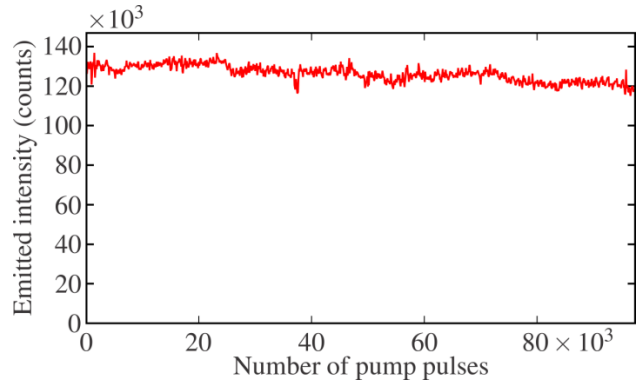

Fig. 7. Photostability curve of a cuboid OMLs (58×58×32μm$^3$) on a PC pedestal fabricated by MPL and pumped just above the threshold.

### 3.3. Lasing thresholds

The lasing thresholds of the cuboid cavities supported either directly by the substrate or by a PC pedestal, that were fabricated by UV and MPL techniques, have been investigated. Because the threshold intensity depends on the resonator shape and decreases as the size increases, the thresholds for cuboid cavities of comparable size were measured. Fig. 8 shows the threshold data for two cuboids fabricated by MPL, with a size of 36×36×35 μm$^3$, one supported by a PC pedestal [cf. Fig. 3(b)], one supported directly on a glass substrate, and a cuboid of size 35×35×31 μm$^3$ fabricated by UV lithography on a Si/SiO$_2$ wafer. Each cavity showed a clear lasing threshold, which is typically one order-of-magnitude lower than for comparable two-dimensional cavities [25]. The lasing threshold of the MPL fabricated cuboid cavity on the PC pedestal is approximately two times smaller than that of the cavity fabricated directly on the glass substrate. This may be associated with a higher refractive index contrast across the interface due to the lower effective index of the PC structure. Clearly, the decoupling of the OML from the substrate by the PC structure leads to a significant reduction of the optical losses from the cavities to the substrate modes. The lasing threshold of cuboid 3D cavities on the PC decoupler is about a factor of four lower than that of the cuboid fabricated by UV lithography on the Si/SiO$_2$ substrate. The thresholds are smaller than those of comparable 2D cavities. Further improvements may be possible by optimizing the surface quality, pedestal geometry, and dye concentration of the cavities. The lasing threshold for the cuboid OMLs attached directly to the substrate was observed to decrease as the cavity height increased from 26 to 46 μm, for a constant edge length (see Fig. 9). Measurements of the absorption coefficient (α) for the PM-597 dye at 0.8 wt% in the MPL cross-linked resin gave a value of α = 2100 cm$^{-1}$, corresponding to a 1/e attenuation of the pump beam over a depth of ~ 5 μm from the top of the cavity, which is substantially smaller than the overall height of the cavity. This suggests that the OML is initially only being effectively pumped in the volumetric region near the top of the cavity (for the current pump configuration). However, the camera image in Fig. 3(b) shows that the lasing modes of thick cavities appear to be localized in the upper part of the cavity. This is consistent with the observation of a 2D diamond PO, independent of the height of the OMLs, over the range investigated. For the shorter OMLs, the losses due to out-coupling of emission into the bulk modes of the glass substrate could result in increased loss. As the OML heights are increased, the distance between the lasing top portion of the cavity and the substrate increases, resulting in a reduction in the lasing threshold. This may be attributed to the larger distance between the pumped top region and the substrate, resulting in lower loss into the substrate. To corroborate this interpretation, a series of OMLs with the same range of heights mentioned above were fabricated on top of BCT-PC pedestals and the lasing threshold energies were determined, see Fig. 10. Contrary to the substrate-supported structures, the lasing thresholds of the PC supported structures were relatively constant, owing to the effective decoupling provided by the PC. Conversely, the thresholds for the shortest cavities placed directly on the substrate were ~ 30 times higher than those on the photonic crystal pedestal.

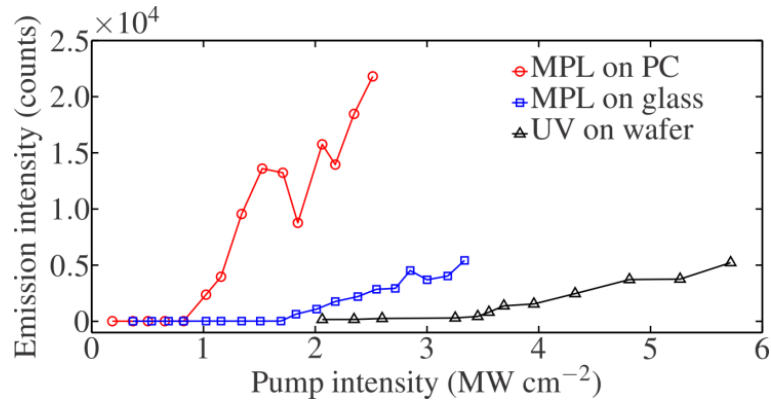

Fig. 8. Lasing threshold data for different OMLs of cuboid shape. The intensity of the laser emission is plotted with respect to the pump intensity for a cavity fabricated by MPL on a PC pedestal (red circles), a cavity fabricated by MPL directly on the glass substrate (blue squares), and a cavity fabricated by UV lithography on a Si/SiO$_2$ wafer (black triangles). The lines are guides to the eye.

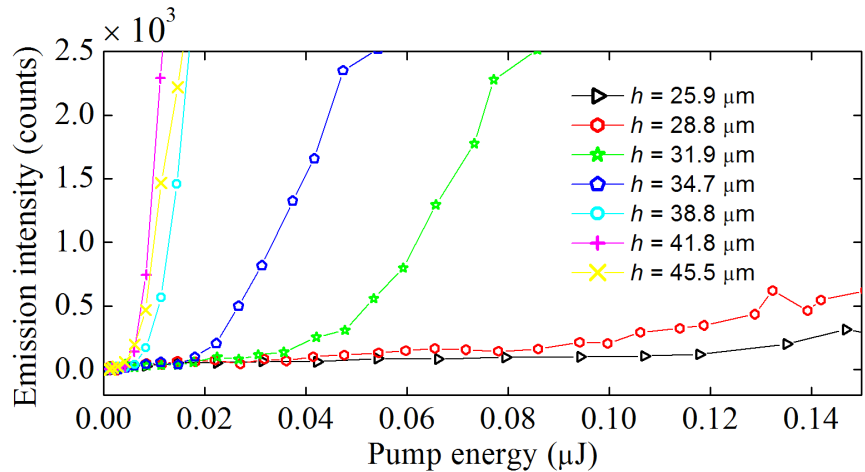

Fig. 9. Lasing threshold curves for OMLs with constant 36 μm edge lengths and varying heights fabricated directly on the substrate. For these measurements, 6 ns, 532 nm laser pulses were used for excitation.

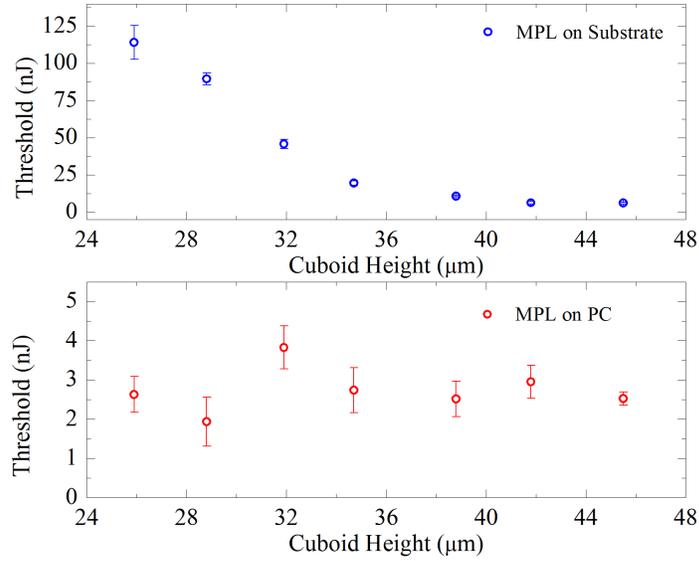

Fig. 10. Lasing threshold data for OMLs with constant 36 µm edge lengths and varying heights fabricated directly on the substrate (blue) or on a photonic crystal pedestal (red). For these measurements, 6 ns, 532 nm laser pulses were used for excitation.

### 3.4. Spectral evolution

As the pump power was increased beyond the lasing threshold, new families of equidistant resonances appeared as seen in the lower part of Fig. 5. The spectrum shown there features a total of four such families. Their FSRs are identical within the precision limits of our experimental setup. Thus, these new families of resonances also appear to be related to the diamond orbit. However, they are based on resonant modes with higher losses. This is demonstrated in Fig. 11 where the amplitudes of four resonances belonging to the four different families are plotted with respect to the pump power. Each of the four curves shows a different threshold, as the different mode families began to lase one after the other when the pump intensity was increased. Theoretical investigations are underway to obtain a detailed understanding of the different types of modes in a cuboid OML and their lasing dynamics.

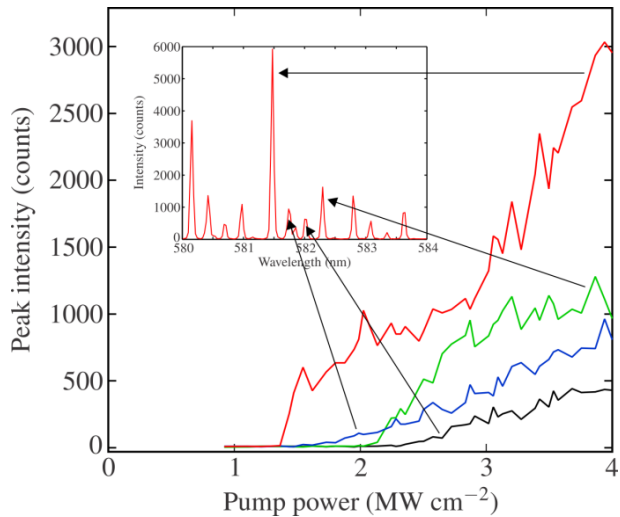

Fig. 11. Threshold curves of resonances belonging to four different families of resonances of a cuboid cavity (58×58×32 µm$^3$) on a PC pedestal fabricated by MPL. The inset shows a portion of the lasing spectrum with the resonances identified.

## 4. Conclusions

We have fabricated true 3D OMLs using UV lithography and MPL techniques. The advantages of the former technique lie in the short time-scale for fabrication and, the large number of structures that can be fabricated at the same time, while the latter technique makes it possible to fabricate arbitrary 3D resonator shapes, including complex pedestal structures. We have investigated the effect of OML dimensions and the use of BCT-PC decoupling structures on the lasing spectra and thresholds. For cuboid OMLs, the lasing spectra feature families of equidistant modes that are apparently localized on a diamond-shaped PO in the horizontal plane, even though the cavities have a similar size in all three dimensions. This is evidenced by the fact that the FSR of the mode families does not depend on the height of the cavities. The confinement of this PO by total internal reflection allows for low lasing thresholds of the associated mode families. Our OMLs based on the PM-597 dye in a crosslinked acrylate polymer show very good photostability with negligible intensity decay over ~100,000 shots, for pumping just over the threshold. The lasing threshold of the OMLs is significantly reduced, by a factor of two to four for cavities supported on BCT-PC structures as compared to OMLs supported directly on the substrate. Under increased pump intensity, new families of higher-loss diamond modes emerge and show different thresholds. We have also observed that as the height of substrate supported OMLs is increased, the lasing threshold is reduced by over an order of magnitude. The emission wavelengths and FSRs can be readily controlled by changing the edge-length of the microresonators. These microresonators can be easily integrated with other optical elements, i.e. waveguides, resonators, and detectors, on a chip or other substrate, and the processing conditions are quite mild compared to undercutting by etching chemistries. The flexibility of this fabrication technique also allows for further studies along the lines of coupled microresonators of various sizes, shapes and spacings for use in sensing, labeling and photonics.

## 5. Acknowledgments

The work at the Georgia Institute of Technology was supported by the U.S. Air Force Office of Scientific Research, BioPAINTS MURI Program (Grant No. FA9550-09-0669). The authors thank Dr. Canek Fuentes Hernandez for assistance with refractive index measurements. The authors thank Gilles Clavier, Rachel Méallet-Renault and Isabelle Leray for fruitful discussions, Maxime Boudreau for ellipsometry measurements, and Valentina Emiliani, Eirini Papagiakounou, and Marc Guillon for preliminary MPL experiments. S.B. gratefully acknowledges funding from the PALM LABEX and the European Union Seventh Framework Programme (FP7/2007-2013) under Grant Agreement No. 246.556.10.